\documentclass[aps,pra,showpacs,twocolumn,amsmath,preprintnumbers,nofootinbib,secnumarabic,a4paper]{revtex4}

\usepackage{epsfig,verbatim,hyperref}
\usepackage{amssymb,amsfonts}
\usepackage{graphicx}

\usepackage[caption=false]{subfig}

\def\art{paper}
\def\jrn#1#2#3#4#5#6{#3 \textbf{#4}, #5 (#6).} \def\boo#1#2#3#4#5#6{\textit{#2} (#3, #4, #5).}                          \def\arx#1#2#3{arXiv:#3.} \def\andd{ and } 

\def\scn#1#2{\section{#1}\lb{#2}} \def\sscn#1#2{\subsection{#1}\lb{#2}} 
\def\eq{Eq.\,} \def\eqs{Eqs.\,} 

\def\bfl{\begin{flushleft}}
\def\efl{\end{flushleft}}
\def\bfr{\begin{flushright}}
\def\efr{\end{flushright}}
\def\bc{\begin{center}}
\def\ec{\end{center}}
\def\be{\begin{equation}}
\def\ee{\end{equation}}
\def\bse{\begin{subequations}}
\def\ese{\end{subequations}}
\def\ba{\begin{eqnarray}}
\def\ea{\end{eqnarray}}
\def\baa#1{\begin{array}{#1}}
\def\eaa{\end{array}}
\def\bw{\begin{widetext}}
\def\ew{\end{widetext}}
\def\nn{\nonumber }
\def\lb#1{\label{#1}}
\def\bit{\begin{itemize}}
\def\eit{\end{itemize}}
\def\bco{}
\def\bcs{\begin{cases}}
\def\ecs{\end{cases}}

\def\schrod{Schr\"odinger}

\def\lan{{\cal L}}
\def\lanp{{\cal V}}

\def\vena{\boldsymbol{\nabla}}

\def\av#1{\langle #1 \rangle}

\def\nc0{\tilde b_0}

\def\icip{\mathbb{U}}
\def\U{{\rm U}}
\def\H{{\rm H}}

\def\vol{{\cal V}}
\def\vol{V}
\def\ello{{\ell_0}}
\def\ncr{{n_c}}
\def\dlin{{\eta}}
\def\ncro{{\dlin_c}}
\def\nrmf{{N}}
\def\wfo{\textbf{a}}

\def\drm{d}
\def\dvol{\drm\vol}
\def\dm{{\bar d}}
\def\ddev{\delta}
\def\phai{A}
\def\phaii{B}

\begin{document}

\preprint{\small Int. J. Mod. Phys. B \textbf{33}, 1950184 (2019)   
\quad 
[\href{https://doi.org/10.1142/S0217979219501844}{DOI: 10.1142/S0217979219501844}]
}

\title{
Temperature-driven dynamics of quantum liquids:
Logarithmic nonlinearity, phase structure and rising force
}

\author{Konstantin G. Zloshchastiev}
\email{https://bit.do/kgz}
\affiliation{Institute of Systems Science, Durban University of Technology, P.O. Box 1334, Durban 4000, South Africa}


\begin{abstract} 
We study a large class of strongly interacting condensate-like materials, which can be characterized
by a normalizable complex-valued function.
A quantum wave equation with logarithmic nonlinearity
is known to describe such systems, at least in a leading-order approximation,
wherein the nonlinear coupling is related to temperature.
This equation can be mapped onto the flow equations of an inviscid barotropic fluid with intrinsic surface tension and capillarity; the fluid 
is shown to have a nontrivial phase structure controlled by its temperature.
It is demonstrated that in the case of a varying nonlinear coupling
an additional force occurs, which is parallel to a gradient of the coupling.
The model predicts that the temperature difference creates
a direction in space in which quantum liquids can flow, even against the force of gravity.
We also present arguments explaining why superfluids;
be it superfluid components of liquified cold gases, or Cooper pairs inside superconductors,
can affect closely positioned acceleration-measuring devices.
\end{abstract}

\date{received: 18 Jan 2019 [EPS], 24 April 2019 [WSPC]} 


\maketitle

\scn{Introduction}{s:in}

Superfluids are a well-known example of quantum materials of a macroscopic size, 
which have been produced in a laboratory \cite{kap38,am38}.
Examples include liquid helium-4 below the 2.17 K (at the normal pressure),
known as  superfluid helium-4 or helium II, and
``bosonized'' fermionic fluid of Cooper pairs of electrons in BCS-type superconductors. 
While in the latter materials superfluidity manifests itself in measurements of mostly electromagnetic properties,
superfluid helium-4 displays visible phenomena of a purely mechanical kind
which are hard to explain using only classical physics \cite{ttbook}.

For example, there is an intriguing phenomenon when superfluid helium-4 creeps up the inside wall of its container and comes down on the outside.
An everyday physics' explanation would be that this is a capillary force of some kind, but this effect 
has certain features, which make it somewhat different from conventional capillarity 
and counterflow phenomena \cite{eh81,gol10arx}.
Firstly, a conventional capillary force is of a conservative type, which means
that it not only elevates a portion of the liquid up but also holds it, 
thus preventing it from dripping, unless additional effects step in. 
On the contrary, superfluid helium flows out of an open container continuously, without
any sign of reversing, or of stopping its flow at some point.
Secondly, a question arises why does the superfluid escape by flowing down on the external
side of a container's wall, instead of flowing down on the internal side, thus returning back to its original point.

These two features indicate that, over and above the conventional capillary 
and counterflow, an 
additional force comes into play,
one
which is directed from the center of the container towards its walls,
thus pushing the fluid not only up but also out.
Such a horizontal force would be mutually compensated inside the fluid's bulk,
but not near container's walls where 
it gets enhanced by molecular bonding with the material the walls are made of.

To summarize these observations,
an additional phenomenon to the conventional capillarity one
should occur,
which is not related to the
interaction of superfluid with the molecular structure of
the container, but it is intrinsic to superfluid itself.
In this \art, we reveal the nature of this additional force.
Moreover, we demonstrate that this is a universal phenomenon
which could occur in a large class of materials; the latter will be specified as well.

\scn{The Model}{s:we}
Following the discovery of Bose-Einstein condensation and related phenomena, 
an understanding came
that quantum Bose liquids are not a mere collection of particles interacting
via some interparticle potential. 
Instead, a new phenomenon occurs:
collective degrees of freedoms emerge, which are no longer
identical to the constituent particles, so that the liquid starts behaving 
as an independent quantum object.
Correspondingly, wavefunctions
related to new degrees of freedom are to be considered fundamental
within the frameworks of the collective approach. 
Even for ground states,
these wavefunctions do not obey the conventional (linear) Schr\"odinger equations,
but some nonlinear quantum
wave equations in which nonlinearities account for many-body effects. 

The fluids with logarithmic nonlinearity, or logarithmic fluids, are one of such models.
The logarithmic fluid approach was originally proposed
in a non-perturbative theory of physical vacuum 
\cite{z10gc,z11appb,az11}. 
Subsequently, the logarithmic models have been generalized 
and applied for a large class of condensate-like materials in which
the
interaction potentials between constituent particles are 
substantially larger than the kinetic energies thereof \cite{z18zna}.
Such materials seem to include not only low-temperature liquids, but also materials 
flowing under special conditions \cite{dmf03,gl08,gl14,z18epl},
which are a subset of materials described by the so-called diffuse interface models
\cite{ds85,amw98}.
Yet another class of systems where logarithmic models might find applications
would be 
neutron stars and quark-gluon plasma objects
where interaction energies dominate kinetic ones
and superfluid components are expected to exist.
Moreover, such models can take into account relativistic and vacuum effects in those
systems, because the local Lorentz symmetry naturally emerges in dynamic equations 
with logarithmic nonlinearity \cite{z11appb,szm16}.

Following the line of reasoning of Ref. \cite{z18zna},
let us consider a fluid, which is connected to a reservoir of infinitely large heat capacity.
Then this fluid can be regarded as
a many-body system, whose particles' interaction energies are larger
than their kinetic ones; 
in what follows
we call such materials condensate-like.
As mentioned above, this class includes not only low-temperature strongly coupled fluids like superfluid helium-4,
but also any matter in which density or geometrical constraints allow the interparticle interaction
dominate over kinetic motion.

From the viewpoint of statistical mechanics,
the microstates are naturally represented 
by a canonical ensemble,
therefore
their probability $P_i$ is given by a standard expression:
$ 
P_i \propto \exp{(-{\cal E}_i/T)} 
$, 
where $T$ is the absolute temperature,
and ${\cal E}_i$ is the energy of an $i$th state;
we use the units where the Boltzmann constant $k_B = 1$.
For above-mentioned materials, one can neglect the kinetic energy,
therefore
the microstate's probability can be approximated by the formula
$ 
P \propto  \exp{(-\U/T)},
$ 
where $\U$ is a characteristic internal potential energy. 

Our second assumption here 
is that the state of such a material can be described by a single complex-valued function
written in the Madelung form \cite{ry99}:
\be\lb{e:fwf}
\Psi = \sqrt n \exp{(i S)}
,
\ee
where $n = n(\textbf{x},t) = |\Psi (\textbf{x},t)|^2$ is particle density,
and 
$S = S(\textbf{x},t)$ is a phase.
The latter is related to the fluid velocity $\textbf{u}$ through the relation $\textbf{u} \propto \vena S$.

This wavefunction naturally obeys a normalization condition
\be\lb{e:norm}
\int_\vol |\Psi|^2 \dvol  
= \nrmf
,\ee 
where $\nrmf$ and $\vol$ are the total amount of particles in the system and its volume, respectively.
This condition imposes conditions 
upon functions $\Psi$,
which reveal the quantum-mechanical nature of our fluid:
a complete set of all normalizable functions  
constitutes a Hilbert space.

Let us assume that the dynamics in such space is of a Hamiltonian flow type
\cite{z18zna}, therefore
$ 
\hat \H |\Psi \rangle = 
\left(
\frac{
\hat{
\textbf{p} 
}^{2}}{2 m} 
+ \hat \U + V_\text{ext}
\right) |\Psi \rangle
$, 
where
$\hat{\textbf{p}} \propto i \vena $
is a generator of spatial translations,
$V_\text{ext} = V_\text{ext} (\textbf{x},t)$ is an external or trapping potential
and $m$ 
is the mass of
a fluid's constituent particle.
Since $|\Psi|^2 \propto P \propto \exp{(-\U/T)}$,
one can choose a potential operator to be
$
\hat \U 
= - \kappa (T - T_c) \ln{(|\Psi|^2/\ncr)}
$,
where $\kappa$ is a  
dimensionless 
proportionality constant, 
and $\ncr$ and $T_c$ are, respectively, values of density 
and temperature at which the statistical effect vanishes.
Here $\hat \U $ must be regarded as a multiplication operator $\hat{O}_f$
satisfying the equation
$\hat{O}_f |\Psi \rangle \propto \ln{(|f|^2)} |\Psi \rangle$,
which is
evaluated on the constraint surface 
$
f - 
\frac{1}{\sqrt\ncr}
\langle \textbf{x} |\Psi \rangle 
= 0
$, further details can be found in \cite{z10gc}.

In summary, the energy conservation law can be written in the position representation 
as a logarithmic \schrod-like equation (LogSE):
\be
i \partial_t \Psi
=
\left[-\frac{\hbar}{2 m} \vena^2
 + \frac{1}{\hbar} V_\text{ext} (\textbf{x},t)
- b 
 \ln\left(|\Psi|^{2}/\ncr\right)
\right]\Psi
,\label{e:o}
\ee
where $\Psi = \Psi (\textbf{x}, t) = \av{\textbf{x} | \Psi}$,
and
$b$ 
is a real-valued parameter of the model.
This equation has long since been known to the physics community \cite{ros68,bb76}; 
but was previously used 
for purposes not related to quantum liquids. 
In our context, this equation describes collective degrees of freedom which emerge
in the above-mentioned condensate-like Bose systems,
due to quantum interactions between original constituent particles,
such as helium atoms.
These collective degrees of freedom thus become new fundamental degrees of freedom \cite{z12eb}.

It also follows from statistical considerations that
the nonlinear coupling must be related to thermodynamic values such as temperature:
\be\lb{e:bcon}
b \sim T \sim T_\Psi
,\ee
where 
$T$ and  $T_\Psi$ are the thermal and quantum-mechanical temperature, respectively,
and symbol ``$\sim$'' means ``a linear function of'' throughout the paper.
The quantum-mechanical temperature is defined as a thermodynamical conjugate
of the quantum information entropy function
$ 
S_\Psi
=
-\int_\vol |\Psi|^2 \ln{\!(|\Psi|^2/\ncr)} \, \dvol
$, 
which was proposed by Everett, and extensively studied by Hirschman, Babenko, Beckner and others 
(some bibliography can
be found in Ref. \cite{z18zna}).
In particular, this entropy function directly emerges from the logarithmic term in \eqref{e:o}
when the latter is averaged using a Hilbert space's inner product \cite{bra91,z10gc,az11}.

From now on, we shall use a more specific version of \eq \eqref{e:bcon}:
\be\lb{e:btemp}
b = - \frac{\kappa}{\hbar} (T - T_c)
,
\ee
where $\kappa$ is assumed to be positive.

\begin{figure}[t]
\centering
\subfloat[$b < 0$]{
  \includegraphics[width=0.49\columnwidth]{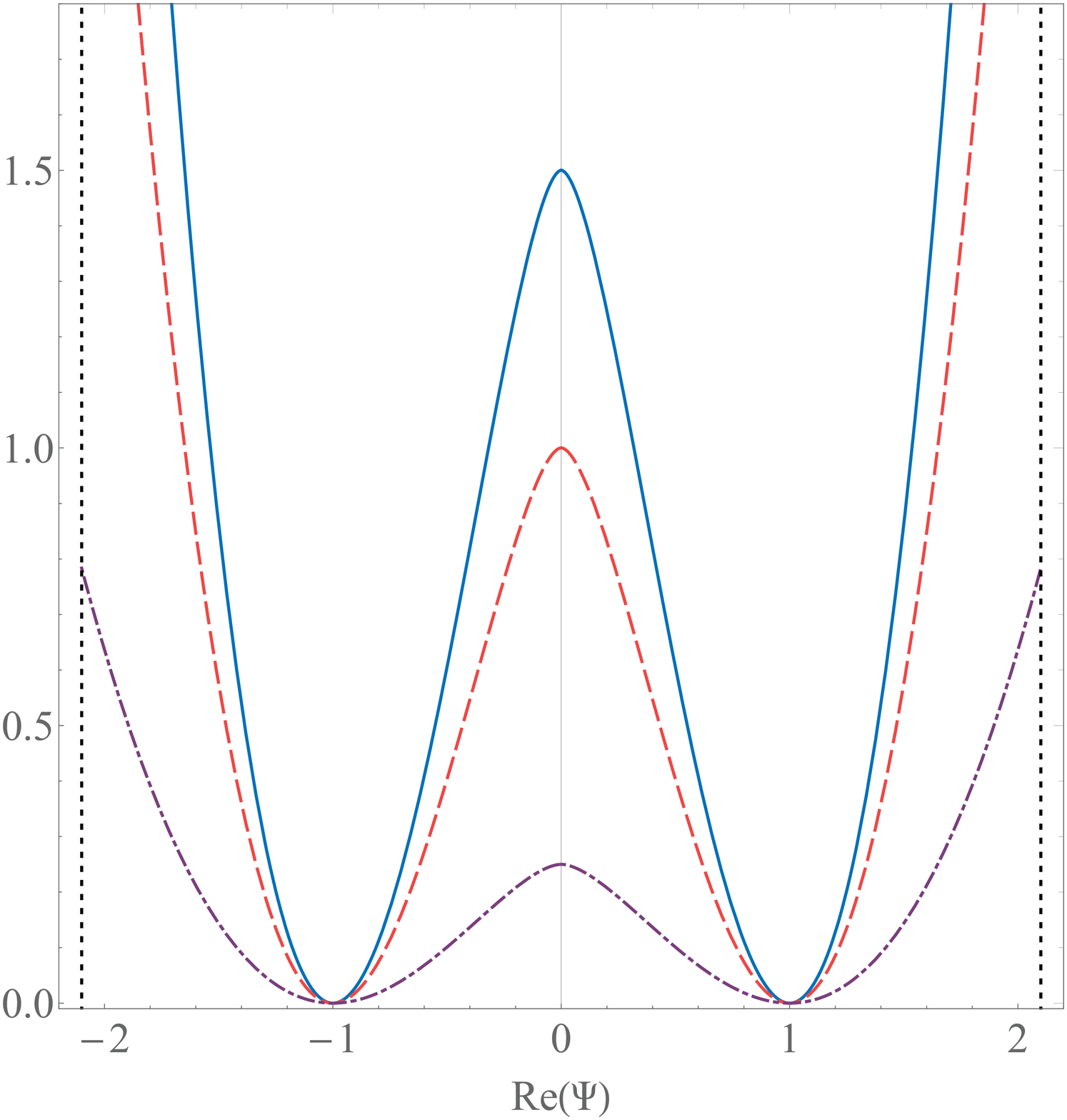}
}
\subfloat[$b > 0$]{
  \includegraphics[width=0.49\columnwidth]{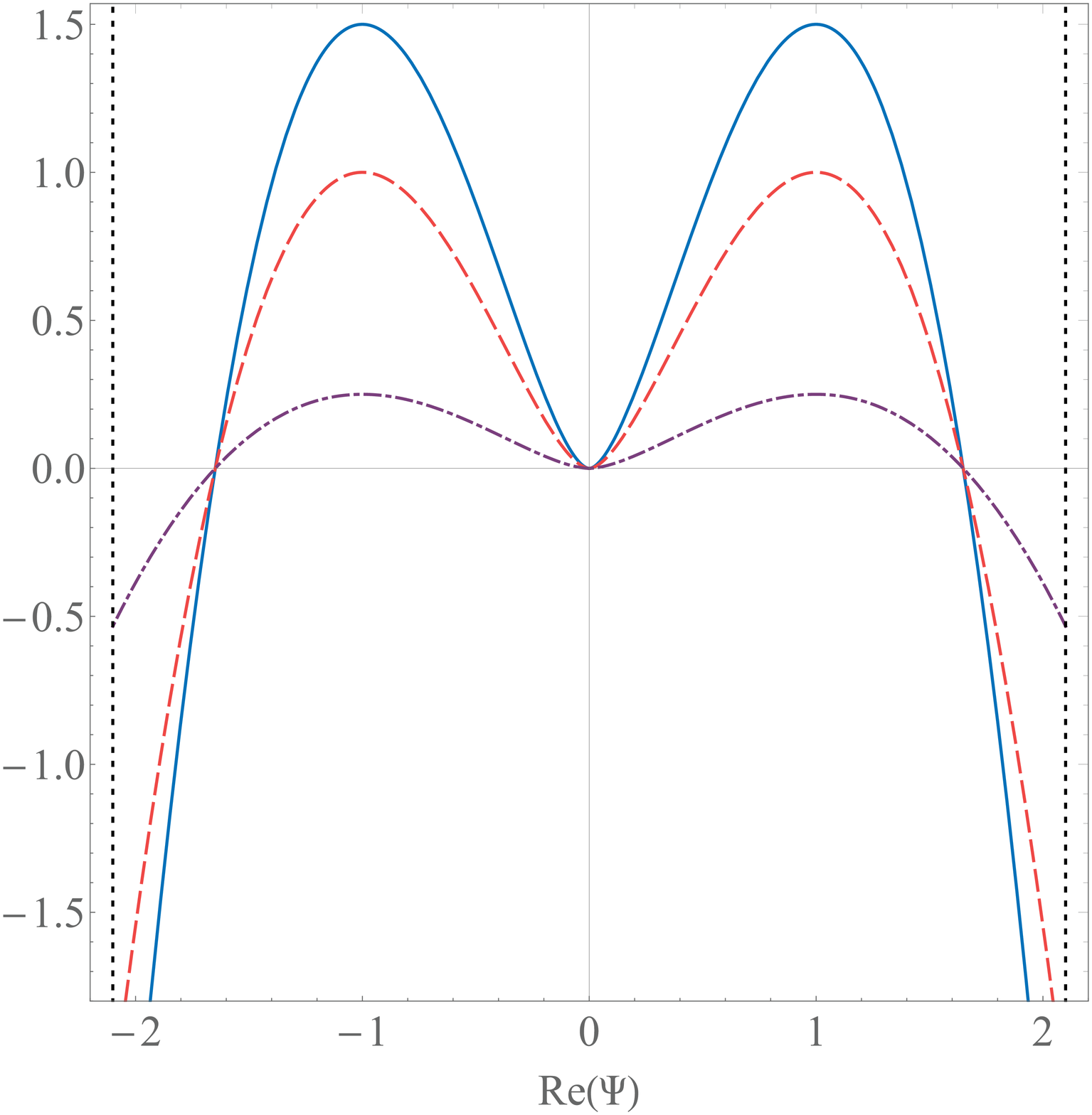}
}
\hspace{0mm}
\caption{Potential 
density $\lanp (|\Psi|^2)$ in units of $\ncr$, versus 
$\text{Re}(\Psi)$ in units of  $|\Psi_e | = \sqrt{\ncr}$, for these values
of $b$: $\pm 3/2$ (solid curve), $\pm 1$ (dashed),  $\pm 1/2$ (dash-dotted),
given in units of inverse time.
Vertical dotted lines represent the condition $|\Psi| \leqslant |\Psi_\text{cut}| < \infty$ which is induced by the 
normalization \eqref{e:norm} acting as a constraint.
}
\label{f:fpotgau}
\end{figure}

\scn{Phase Structure}{s:sb}
Let us discuss the phase structure of a generic logarithmic fluid described by \eqs
\eqref{e:o} and \eqref{e:norm}.
For simplicity,
let us neglect the boundary effects and consider a trapless flow, \textit{i.e.}, $V_\text{ext} \equiv 0$.
By analogy with Refs. \cite{z11appb,z18zna},
by substituting \eq \eqref{e:fwf} into \eqref{e:o} 
and assuming $\textbf{u} = (\hbar/m) \vena S$ and $\rho = m\, n$,
one recovers the
hydrodynamic equations for mass and momentum conservation describing a
two-phase inviscid fluid with capillarity and internal surface tension:
\ba
&&
\partial_t\rho 
+ \vena\cdot(\rho\,\textbf{u})
= 0
,\lb{e:floma}\\&&
\frac{D \textbf{u}}{D t}
\equiv
\partial_t \textbf{u}
+
\textbf{u} \cdot\vena \textbf{u}
=
\frac{1}{\rho} \vena\cdot \mathbb{T}
,
\lb{e:flomo}
\ea
with the stress tensor $\mathbb{T}$ being of the Korteweg form 
with capillarity \cite{ds85}:
\be\lb{e:stko}
\mathbb{T} 
=
-\frac{\hbar^2}{4 m^2 \rho} \vena\rho \otimes \vena\rho 
- \tilde p \, 
\mathbb{I}
,
\ee
where $\mathbb{I}$ is the identity matrix, 
$
\tilde p 
=
p (\rho) - (\hbar/2 m)^2  \vena^2\rho 
$, 
and $p (\rho) = - (\hbar b/ m) \rho$ is a barotropic equation of state for the pressure $p$.

On the other hand,
\eq \eqref{e:o} can be derived by minimizing
an action functional with the following Galilean-invariant Lagrangian density:
\be\lb{e:ftlan}
\lan/\hbar
= 
\frac{i}{2}(\Psi \partial_t\Psi^* - \Psi^*\partial_t\Psi)+
\frac{\hbar}{2 m}
|\vena \Psi|^2
+
\lanp (|\Psi|^2)
,
\ee
with the potential density 
\be\lb{e:ftpot}
\lanp (n) = -
b n
\left[
\ln{(n/\ncr)} -1
\right]
+ \lanp_0
,
\ee
where $\lanp_0 = \lanp (0) $ is a potential's shift constant; its value is chosen as to ensure
that the function \eqref{e:ftpot} always
vanishes at its local minima:
$\lanp_0 =  \ncr (|b|-b)/2$.
For complex $\Psi$'s,
this potential has nontrivial local extrema at $|\Psi_e | = \sqrt{\ncr}$.
Correspondingly,
they become minima (maxima) if $b$ is negative (positive),
as shown in Fig. \ref{f:fpotgau}.

These features indicate a symmetry breaking/restoration,
and the existence of at least two phases in the fluid.
Let us separately study what happens in each of these phases.

\sscn{Phase \phai}{s:phai} 
When the liquid temperature is above
the critical value, $T > T_c$, then the nonlinear coupling $b$ is negative,
according to \eqref{e:btemp}.
The potential density $\lanp$ acquires the conventional Mexican hat shape
if plotted in a space of real and imaginary parts of $\Psi$,
with local degenerate minima located at $|\Psi_e |$, 
see fig. \ref{f:fpotgau}a. 
Therefore, topologically nontrivial solutions exist which interpolate between the local
minima \cite{z11appb}. 

Due to separability of \eq \eqref{e:o} in Cartesian coordinates (which makes it different
from other nonlinear wave equations),
one can impose the ansatz
$ 
\Psi (\textbf{x}, t) = \prod\limits_{j=1}^\dm \psi_j (x_j, t)
$, 
where 
$ 
\int_{X_j} |\psi_j|^2 \drm x_j  = 
\int_{X_j} \dlin_j \, \drm x_j = 
\nrmf^{1/\dm}
$, 
and $\dlin_j = |\psi_j|^2$ and $X_j$ 
are fluid's linear density and extent along a $j$th coordinate, respectively.

Then \eq \eqref{e:o} splits into $\dm$ 
one-dimensional equations:
\be\lb{e:osep}
i \partial_t \psi_j
=
\left[-\frac{\hbar}{2 m} \partial_{x_j x_j}^2
+ |b| 
 \ln\left(|\psi_j|^{2}/\ncro \right)
\right]\psi_j
,
\ee
where $\ncro = \ncr^{1/\dm}$, and $\partial_{x_j}$ is a spatial derivative with respect to $j$th coordinate;
a parametric change $b \mapsto - |b|$ is introduced for uniform notations in subsequent formulae.
Therefore, in this case we deal with only one 1D differential equation,
so the index $j$ can be omitted for brevity.

\begin{figure}[t]
\centering
\subfloat[LogSE kink]{
  \includegraphics[width=0.49\columnwidth]{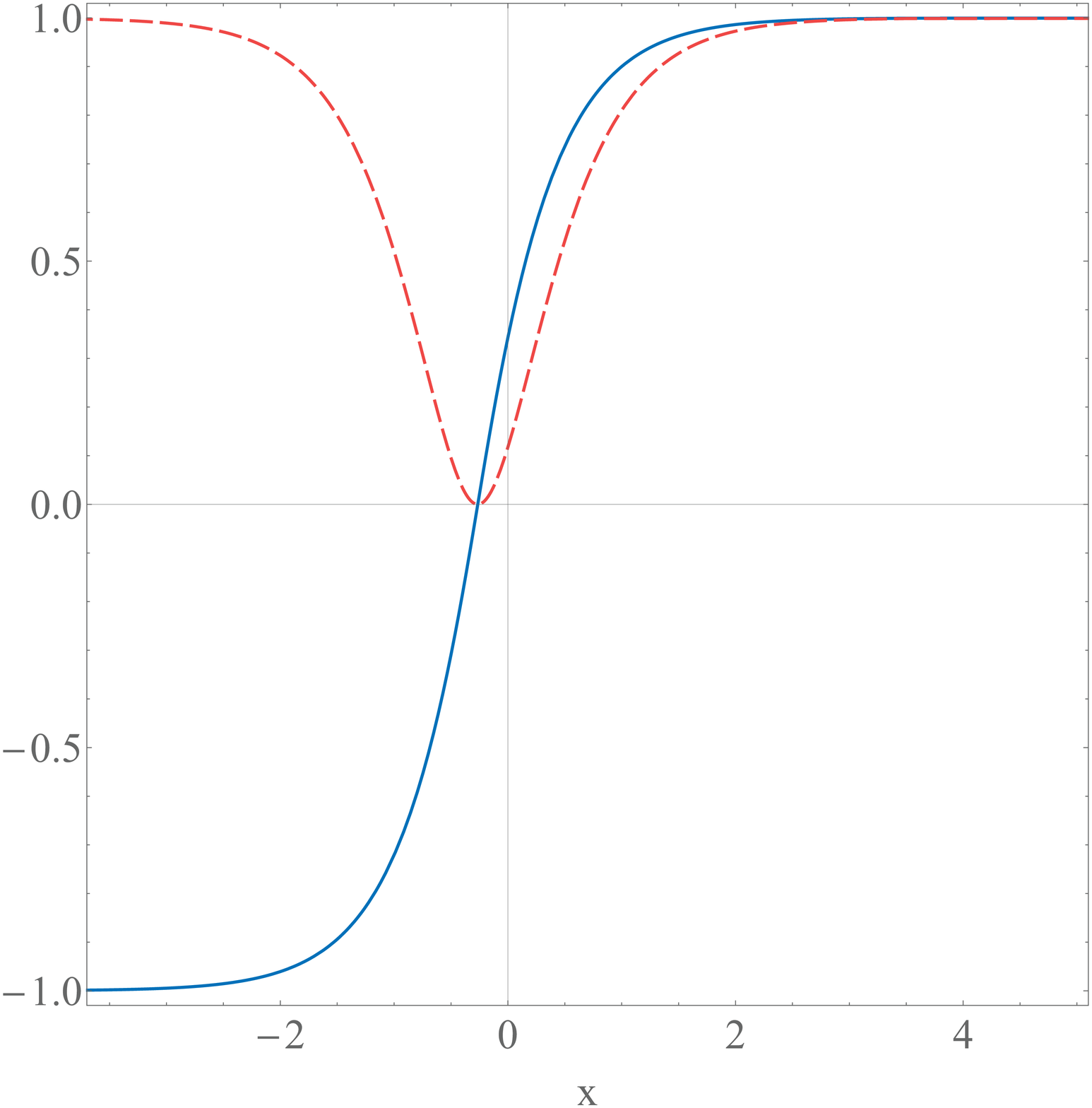}
}
\subfloat[LogSE anti-kink]{
  \includegraphics[width=0.49\columnwidth]{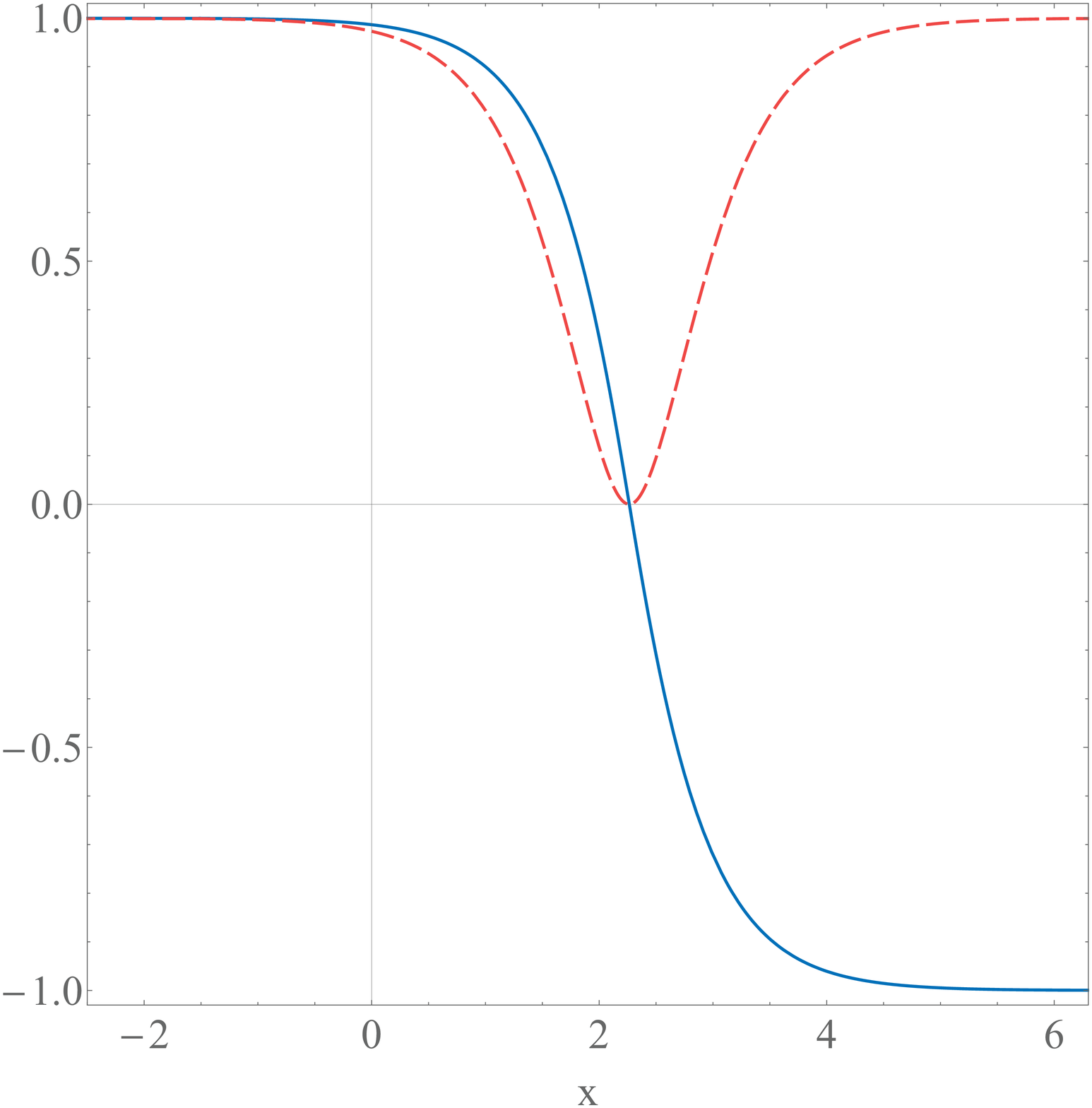}
}
\hspace{0mm}
\caption{Profiles of the topological solitons $\psi_\pm$ (in units of $\sqrt{\ncro}$, solid curves) 
and corresponding
densities $\dlin_\pm$ (in units of $\ncro$, dashed curves), versus the Cartesian coordinate $x$ (in units of  $\ell$);
the upper (lower) sign corresponds to the kink (anti-kink) solution.
For computations we used 
the boundary condition 
$\psi(\infty) =  \pm\sqrt{\ncro}$.
}
\label{f:kink}
\end{figure}

Furthermore, 
our solitons are expected to
saturate the Bogomolny-Prasad-Sommerfield (BPS) bound \cite{Rajaraman:1982is},
\textit{i.e.}, obey
an 
equation
$ 
d \psi/ d x
= \pm\sqrt{2 \tilde U (\psi)}
,
$ 
where $\tilde U (\psi) = 
(2 m |b|/\hbar)
\left\{
|\psi|^2
\left[
\ln{(|\psi|^2/\ncro)} -1
\right]
+ 
\ncro
\right\}
$ is a soliton particle potential,
$x \in \{ x_1,...,x_\dm \}$,
and the sign $\pm$ refers to the soliton and anti-soliton branches.
While the analytical solution for this case is unknown, numerical studies 
reveal
the existence of topological solitons of a kink type, see fig. \ref{f:kink}.
Despite these solutions not being ground-state ones, their longer lifetime is guaranteed by a topological charge,
$Q = \ncro^{-1/2}[\psi(+\infty) - \psi(-\infty)]$ which is nonzero for kinks.
This separates the topologically nontrivial sector our solitons belong to from the sector with $Q \equiv 0$.
The latter has a trivial topology and contains all states with $|\psi |  = \text{const}$, including the ground state.
Though, in real physical conditions, \textit{i.e.}, in the presence of quantum fluctuations and environment's 
influence, these metastable states would eventually decay by emitting waves,
which corresponds to a transition to a gas phase.

In a single-soliton setup, these solutions extend for the whole spatial axis, but in a real fluid
kinks match with antikinks 
with periods of order $\ell$.
This matching can be visualized by repeating the merged fig. \ref{f:kink} along all spatial directions.
According to this figure, the kink soliton's density increases from its center of mass outwards. 

Therefore, a logarithmic fluid in this phase tends to nucleate bubbles with a characteristic size $\ell$,
thus forming a foam-like structure, which precedes 
the process of releasing of vapor or previously dissolved gas, \textit{i.e.}, boiling.
If the temperature continues to grow, then boiling eventually occurs, thus indicating a standard vapor-liquid phase transition whose properties can be studied using  conventional kinetic theory.

\sscn{Phase \phaii}{s:phaii} 
When temperature drops below
a critical value, $T < T_c$, then the nonlinear coupling $b$ turns positive.
Therefore,
the potential density $\lanp$ acquires an upside-down Mexican hat shape
if plotted in a space of real and imaginary parts of $\Psi$,
with local degenerate maxima located at $|\Psi_e | = \sqrt{\ncr}$,
see fig. \ref{f:fpotgau}b.

\begin{figure}[t]
\centering
\subfloat[LogSE gausson]{
  \includegraphics[width=0.49\columnwidth]{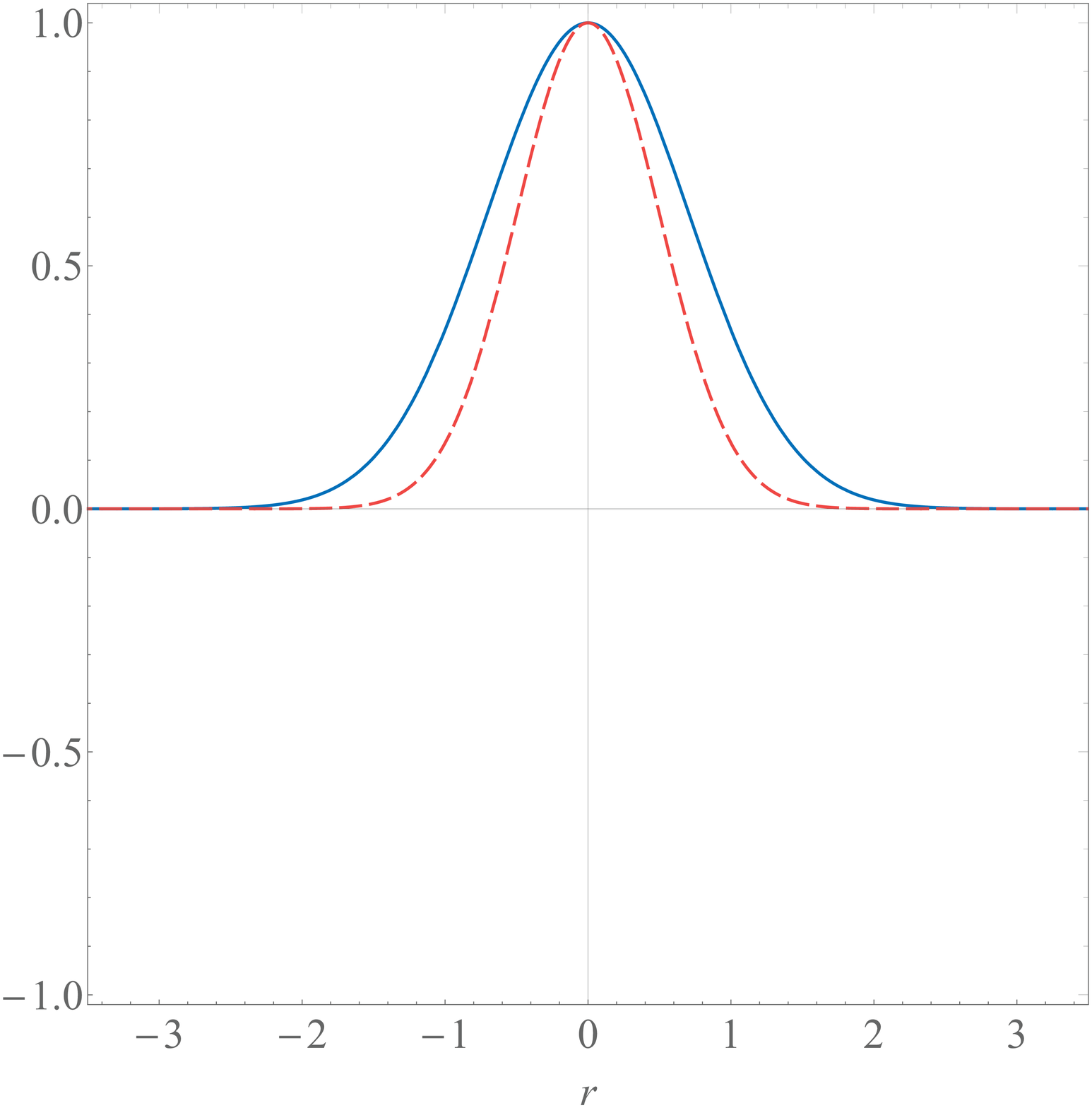}
}
\subfloat[LogSE anti-gausson]{
  \includegraphics[width=0.49\columnwidth]{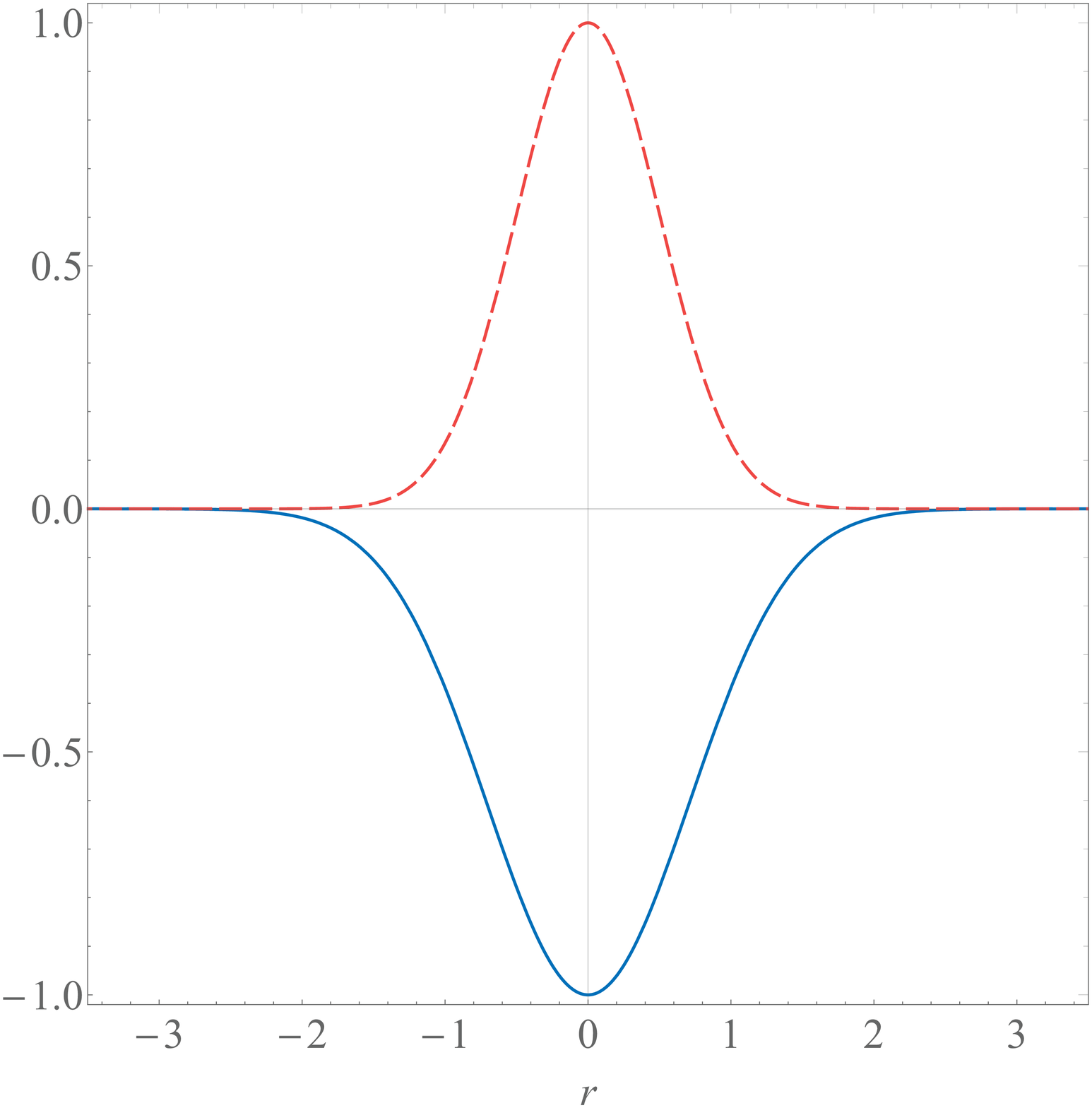}
}
\hspace{0mm}
\caption{Profiles of the Gaussian solitary waves $\Psi_{(g)} (\textbf{x})$, in units of $\sqrt{\tilde n}$ (solid curves), 
and corresponding densities $n_g (\textbf{x})$, in units of $ \tilde n$ (dashed curves), versus the Cartesian axis projected coordinate $r$,
in units of  $\ell$.  
}
\label{f:gauss}
\end{figure}

For a positive $b$, the ground state solution of \eq \eqref{e:o} can be found analytically \cite{ros68,bb76}.
It can be
written
in a form 
of a Gaussian packet modulated by a plane wave:
\be
\Psi_{(g)} (\textbf{x}, t) = 
\pm 
\sqrt{n_g (\textbf{x})}
\exp{\left(- i \omega_g t
\right)} 
,
\ee
where 
\ba
n_g (\textbf{x})
&=&
\tilde n
\exp{\left[- \frac{(\textbf{x}- \textbf{x}_0)^2}{\ell^2} \right]}
,\lb{e:gauss}\\
\omega_g
&=&
b 
\left[
\dm - 
\ln{
 \left(
        \tilde n/\ncr
 \right)}
\right]
,
\ea
where
$\dm = 3$ is a number of spatial dimensions,
and
$\ell = \sqrt{\hbar/2 m |b|}$,
$\tilde n = \nrmf/\tilde\vol$
and
$
\tilde\vol= 
\pi^{\dm/2} \ell^\dm
$
are, respectively, a Gaussian width of the solution, its 
density's peak value and its effective volume.
The solutions' profiles are plotted in fig. \ref{f:gauss}.

The stability of such solutions was confirmed in \cite{az11,bo15,z17zna},
therefore in this phase our fluid tends to
fragment into a 
cluster
of Gaussian density inhomogeneities, each
described by \eq \eqref{e:gauss},
referred to here as fluid elements.
This structure can be visualized by repeating the merged fig. \ref{f:gauss} in all directions.

One can treat these elements 
as point particles of mass $M$ (in general, $M \not= m$),
while effectively encoding their nonzero
size in the nonlocal two-body interaction potential 
$\icip \left( |\textbf{x} - \textbf{x}' | \right)$.
In the leading-order approximation,
it can be written as
\be\lb{e:intcel}
\icip (r) \approx
\frac{U_0}{\ell} (r - \ello) \, 
\exp{\left[-(r/\ell)^2\right]}
,
\ee
where 
$r = |\textbf{x} - \textbf{x}' |$,
$U_0$ is
the proportionality factor,
and
$\ell_0 
= \ell
\left[
1/2 + 1/(\text{e} \sqrt{\pi}\, \text{erf}(1))
\right] + \delta \approx 0.75 \ell + \delta$,
where $\delta$ being a small parameter responsible
for size fluctuations,
$\delta \ll \ell$, see Ref. \cite{z12eb} for further details.

\scn{Empirical Evidence}{s:ob}
In this
section, we shall give a brief overview of previous results on logarithmic models' usage
for quantum liquids and dense Bose-Einstein condensates; 
an emphasis will be given on the experimental evidence existing to date,
which supports applicability
of these models in the theory of superfluidity of liquid
helium.

To begin with, if one assumes 
that
\be
T_c \lessapprox T_\lambda \approx 2.17 \, \text{K}
\ee
at normal atmospheric pressure,
then the above-mentioned phase structure 
explains why liquid helium-4 is actively boiling above the $\lambda$-point  but remains totally still just under,
thus contributing to the effect caused by a jump of heat conductivity.

The logarithmic model analytically describes not only the superfluidity of helium-4,
which will be
discussed below, but also the phase transition between the foam-like (pre-boiling) and gausson (still) phases,
referred to above as phases \phai\, and \phaii, respectively.
Strictly speaking,
this is an effect, which is separate from the inviscid flow phenomenon \textit{per se}.
It merely states that the formation of bubbles in quantum Bose liquids is a topological effect, which is
allowed when the logarithmic nonlinear coupling $b$ is negative,
and is forbidden otherwise.   
In other words, the topological structure of liquid helium changes 
drastically when
crossing the $\lambda$-point. 
Further above this point, bubble formation eventually leads to the evaporation of the helium,
so that the original requirements for a condensate-like material 
are no longer valid; hence one has to use the methods and notions of 
the standard kinetic theory.

Furthermore, the above-mentioned Gaussian droplet-like phase
is important for understanding the
behavior of liquid helium-4 below the $\lambda$-point, historically referred as the helium II phase.
In Ref. \cite{z12eb}, it was shown
that 
this model is very instrumental for describing the
superfluid component of helium-4.

For instance,
the Landau's form of an excitation spectrum, \textit{i.e.}, the one 
which contains a local maximum (``maxon'') followed by a local minimum (``roton''), is a well-known sufficient condition
for superfluidity to occur. 
For helium II, the experimental data coming from neutron scattering put a stringent constraint on a position of both  these local extrema, let alone that a candidate model is also expected to reproduce the structure factor curve (coming from X-ray scattering experiments) and speed of sound.
Nevertheless,
the logarithmic fluid model managed to analytically reproduce with high accuracy
three main 
observable facts of this liquid -- Landau spectrum of excitations,
structure factor and speed of sound -- while using only one non-scale parameter to fit
the  excitation spectrum's experimental data.

The resulting theory turned out to be a two-scale model, 
where the shorter-scale part invoked the wave equation \eqref{e:o}, with positive nonlinear coupling, for describing the ground state
of the superfluid component of liquid helium and the
formation 
of the effective degrees of freedom called Gaussian fluid elements or parcels. 
The longer-scale theory picks up from there, deriving the potential \eqref{e:intcel} 
as a leading order approximation for interaction between Gaussian elements.
It should be emphasized that, while for describing the Landau spectrum the potential \eqref{e:intcel} is formally
sufficient, for describing the structure factor and speed of sound, one needs both 
this potential and the  
wave equation \eqref{e:o}.
This confirms the logarithmic model's applicability for the superfluid helium phenomenon at different levels.

For superfluid helium-4, 
the numerical values of the parameters of the two-scale model are
\ba
&&
b  \approx 0.45\, \Delta/\hbar ,\
m = m_{\text{He}} \approx 6.64 \times 10^{-24} \text{g}, \nn\\&&
\ell \approx 1.25 \ \text{\AA},\
U_0  \approx 69 \, \Delta,\
M/m \approx 1.23,
\ea
where $\Delta \approx 8.65$ K is a ``roton'' energy gap for helium-4;
one can impose that $n_c \approx 1/\ell^3$ here.

For the phase \phai, at 
$T > T_c$, the analytical form of interaction potential between 
the fluid elements is currently unknown.
However, one can naturally expect that such a potential 
belongs to a class of hard-sphere potentials which are 
popular in the theory of cold quantum liquids \cite{psb18}.
The common feature of such potentials is that 
they have a global maximum of repulsive energy at a sphere surface's radius, and abruptly become zero
when moving away from it outwards.
This 
qualitatively resembles behaviour of bubbles which nucleate in superfluid and
compose the phase \phai, as was shown before.

On formal grounds, the logarithmic fluid in this phase is still having a zero viscosity,
according to \eq \eqref{e:flomo}.
However, this is different from phase \phaii, because the topological solitons representing the phase-\phai\, bubbles are not ground states of \eq \eqref{e:o}.
Instead, they are excited states, although the metastable (or long-lived) ones, due to topological arguments discussed above.
This means that phase \phai\,
describes a critical superfluid which is just about to boil, \textit{i.e.},
it is an intermediate phase between the phase \phaii\, and the phase of active boiling and vaporizing.  

Obviously,
due to thermal fluctuations,
logarithmic superfluid in the vicinity of the critical temperature $T_c$
must be 
a mixture of both phases.
Therefore, the statistical mechanical properties
of fluid elements of such liquid must be governed by a potential, which is 
a combination of the potential \eqref{e:intcel} and a potential of a hard-sphere type.
 
To conclude, an analytical simplicity of a model together with its good agreement with experiment,
clearly indicates that the quantum fluid with logarithmic nonlinearity
should be a main component of models of liquid helium at low temperatures.
This, of course, does not preclude from using
other ingredients,
such as the Gross-Pitaevskii condensate (contact two-body interactions, pertinent to weakly-interacting cold gases) 
and sextic fluid (three-body interactions and Efimov states),
which can provide higher-order corrections.

\scn{Additional Force}{s:ada}
Until now we have been dealing with an isothermal logarithmic fluid.
If, assuming the liquid to be in the phase \phaii, 
one substitutes a solution 
\eqref{e:gauss} into a right-hand side of \eqref{e:flomo}
then it turns out to be zero, 
which means that the material derivative of the flow's averaged velocity vanishes.
This indicates that inside the ground-state superfluid flow of constant temperature
all forces are compensated.
What would happen if one introduced a small temperature gradient into the system while keeping its
temperature below $T_c$?

The simplest way to do so is to recall \eq \eqref{e:btemp}
and assume that the nonlinear coupling is no longer
a constant but a function 
\be
b = b (\textbf{x}, t) \
\Rightarrow \
T = T (\textbf{x}, t)
.
\ee
Then the mass conservation equation \eqref{e:floma} remains intact,
but the momentum equation changes 
from \eq \eqref{e:flomo} to this one:
\ba
\frac{D \textbf{u}}{D t}
=
\frac{1}{\rho} \vena\cdot \mathbb{T}
+
\wfo
,
\lb{e:flomov}
\ea
where the additional acceleration term
reads:
\be\lb{e:wfo}
\wfo = \frac{\hbar}{m}  \ln{\!(\rho/\rho_c)}\, \vena b (\textbf{x}, t)
,
\ee
where $\rho_c = m \ncr$.
This acceleration looks as if it were caused by
an external force $\textbf{f} = \rho\, \wfo$
acting on a fluid parcel,
which is parallel to the gradient of $b (\textbf{x}, t)$,
whence
$\textbf{f}$ must be parallel to the temperature gradient,
according to \eq \eqref{e:btemp}.

Let us consider small variations of temperature
which do not violate the condition $T < T_c$.
Using relation \eqref{e:btemp}, formula 
\eqref{e:wfo}
can be rewritten as
\be\lb{e:wfo2}
\wfo 
= 
- \frac{\kappa}{m}   \ln{\!(\rho/\rho_c)} \vena T 
\approx
- \frac{\kappa}{m}   \ddev \, \vena T  
,
\ee
where 
$\ddev = (n - \ncr)/\ncr = (\rho - \rho_c)/\rho_c$ is a relative deviation of density from its critical value;
the approximation in this equation is valid only if $\ddev $ is small.
Since $\kappa$ was chosen to be positive by construction, vector $\wfo$
is parallel to the temperature gradient
if $\ddev < 0$, and anti-parallel otherwise.

\begin{figure}[h]
\epsfig{figure=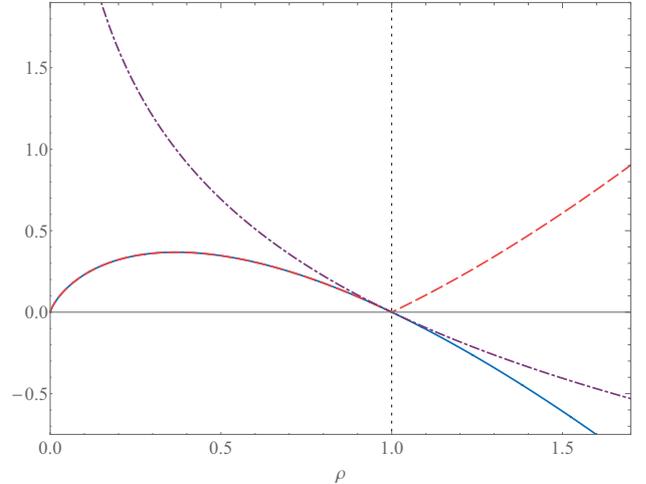,width=  0.95\columnwidth}
\caption{Profiles of dimensionless magnitude functions from \eq \eqref{e:magfa}, 
versus density in units of  $\rho_c$:
$K_f$ (solid curve), $|K_f|$ (dashed curve) and $K_a$ (dash-dotted curve).
}
\label{f:signmag}
\end{figure}

Let us study a behaviour of that part of force and acceleration, which is related to density only,
by re-writing the additional force and acceleration as
\be\lb{e:magfa}
\textbf{f} = 
\frac{\kappa}{m} \rho_c
K_f 
\vena T
,\ \
\wfo = 
\frac{\kappa}{m}
K_a
\vena T
,
\ee
where 
$K_f = -(\rho/\rho_c)   \ln{\!(\rho/\rho_c)}$
and
$K_a = -\ln{\!(\rho/\rho_c)}$
are dimensionless functions of density.
Notice that at $\rho\to 0$, vector $\wfo$ does not
vanish (moreover, it actually diverges),
but the force nevertheless tends to zero.
For a fixed value of temperature gradient,
it is the function $K_f$ which determines a 
parallel or anti-parallel direction of the force acting on our system,
see fig. \ref{f:signmag}.

One can see that the force changes its direction
when density goes across the critical value $\rho_c$.
Moreover, at densities below $\rho_c$
the density-related magnitude of the force has a local maximum,
at $\rho \to \rho_c/\text e$.
This can be used for establishing a value of $\rho_c$ empirically,
without placing superfluid into the critical regime,
as well as for searching the conditions under which this force is most visible.

Some concluding remarks are in order here.
Firstly, $\rho_c$ or $\ncr$ is not necessarily the density of all fluid at $T = T_c$.
Due to quantum fluctuations,
the actual density can be different from this value, let alone that it can be a function of space and time in general.
Therefore, $\ncr$ must be treated as a parameter of the logarithmic
model,
which marks a critical density value at which the logarithmic nonlinearity 
flips its sign. 
Secondly,
the presented analysis is based on a simplified setup of the irrotational superfluid being
free of any external
forces and containers
(logarithmic fluids in certain types of external potential traps, such as harmonic or Coulomb  
potentials, were discussed in, \textit{e.g.}, refs. \cite{bb76,bo15,ss18}).
If one considers other physical setups then both the value and direction 
of this additional acceleration can be different from formula \eqref{e:wfo2},
but the fundamental reasons for the effect remain intact.

\scn{Conclusion}{s:con}
In this \art, we 
used statistical mechanical arguments
to derive an evolution equation for quantum liquid in a condensate-like state, i.e.,
when average kinetic energies of constituent particles are small compared to the interaction potentials.
This equation was shown to contain logarithmic nonlinearity with respect to the condensate function.
In the classical hydrodynamics context, this equation can be mapped on the flow of the Korteweg-type materials and diffuse interface models with capillarity and intrinsic surface tension.

The corresponding nonlinear (logarithmic) coupling is shown to be a linear function of the liquid's temperature.
It can take
both positive and negative values depending on whether the liquid is, respectively, below or above a critical value.
Therefore, the nonlinear coupling's sign marks two different phases the liquid could be in.

The higher-temperature phase, or phase \phai~in our notations, 
is characterized by metastable (long-lived) states described by
topological solitons with a kink profile,
which 
belong to a topological 
sector different from the one with the trivial solution.
Such solitons can be naturally interpreted as the nucleating
bubbles.
The foam-like structure which occurs facilitates release of vapor or any gas previously dissolved in the fluid,
which is a phase preceding the process of boiling.

The lower-temperature phase, or phase \phaii, 
is characterized by fluid being in a ground state described by the well-known 
solution whose density has a Gaussian profile.
In a multi-solution picture,
this gives rise to a fluid consisting of 
cell-like inhomogeneities.
For such a cellular structure,
both
an effective interaction potential and many-body Hamiltonian were presented.

It was shown that the logarithmic fluid 
can be used to model 
superfluid components of quantum Bose liquids,
such as liquid helium-4.
The corresponding experimental evidence was discussed. 

Furthermore,
it turns out that if one introduces a temperature gradient into the logarithmic fluid
then the additional force per element arises.
This leads to our model being able to predict or explain a number of mechanical phenomena
occurring 
in quantum liquids.

For instance,
in the case of superfluid being confined inside a container at a homogeneous gravitational field,
at least 
two temperature gradients usually exist.
The first gradient, a horizontal one, occurs when the temperature of the superfluid at its center 
is slightly
different from the temperature on its border adjacent to the container's walls.
The second gradient, a vertical one, occurs along the container's walls,
because the part of a usually non-full container which is in direct contact with superfluid
has a slightly different temperature than its rim.

It has been shown that logarithmic superfluid tends to flow along the horizontal gradient of temperature
away from a container's center,
as well as
along the vertical gradient of temperature between the near-wall parts of superfluid
and the container's rim. 
This does resemble the known behaviour of liquid helium-4 below the $\lambda$-point:
it creeps up to the highest point, which corresponds to an equilibrium between the gravity force, additional force $\textbf{f}$  and molecular bonding with container's walls,
but does not stay there, nor does it return to the inside of its vessel, but instead flows away on the outside wall.

Yet another prediction which follows from our model is that quantum Bose liquids,
be it superfluid components of liquified cold gases or Cooper pairs inside superconductors,
should affect closely positioned
acceleration-measuring devices, such as gyroscopes, 
which resembles the outcomes of a frame-dragging phenomenon in general relativity
(see also remarks about relativistic effects in Sec. \ref{s:we}).
This becomes possible due to superfluid being a macroscopical yet quantum object, therefore 
it is not sharply localized within its
classical borders.
In our model, a wavefunction has a Gaussian ``tail'', whereby it can interact
with surrounding objects, while the strength of such interaction is determined by the formula \eqref{e:wfo2}
or version thereof.

The magnitude of the above-mentioned effects is determined by a product of temperature gradient
and density difference $\ddev$, on top of a magnitude of the numerical factor $\kappa/m$.
Both the temperature gradient and density variations are small in a typical experimental setup \cite{eh81,gol10arx},
which should result in a statistical smallness of the overall effect in the experiments
\textit{akin} to those described in refs. \cite{chi82,tfc90,tps09,tp10}.
Therefore,
in order to achieve a viable effect,
it is important to have substantial control over these two parameters.

On the theoretical side, future studies will be devoted to extended models of trapped quantum liquids,
which would take into account boundary effects from vessels, rotational dynamics and external potentials, such 
as harmonic traps or homogeneous gravitational field;
those might provide corrections or modifications of the formula \eqref{e:wfo2}.

~\\ \textbf{Acknowledgments}\\
I am grateful to participants of the 8th International Conference on Applied Physics and Mathematics ICAPM-2018 in Phuket, Thailand (27-29 January 2018) \cite{z18cs1}, and XXIII Fluid Mechanics Conference KKMP-2018 in Zawiercie, Poland (9-12 September 2018) \cite{z18cs2}
for stimulating discussions and remarks.
This work is based on the research supported 
by the National Research Foundation of South Africa 
(under Grants Nos.
95965,
98892 and 
117769),
as well as by the Department of Higher Education and Training of South Africa.
Proofreading of the
manuscript by P. Stannard is greatly appreciated.



\end{document}